\documentclass
[superscriptaddress,secnumarabic,amssymb,amsmath,nobibnotes,aps,prd,showkeys,showpacs,nofootinbib,onecolumn,notitlepage]{revtex4}%
\usepackage{setspace}
\usepackage{color}
\usepackage{amsmath}
\usepackage{amsfonts}
\usepackage{verbatim}
\usepackage{amssymb}
\usepackage{graphicx,bm}
\usepackage{graphicx}
\usepackage{bbm}
\usepackage{amsmath}
\usepackage{amssymb}%
\providecommand{\U}[1]{\protect\rule{.1in}{.1in}}
\newcommand{\be}{\begin{equation}}
\newcommand{\ee}{\end{equation}}

\newcommand{\mincir}{\raise
-3.truept\hbox{\rlap{\hbox{$\sim$}}\raise4.truept\hbox{$<$}\ }}
\newcommand{\magcir}{\raise
-3.truept\hbox{\rlap{\hbox{$\sim$}}\raise4.truept\hbox{$>$}\ }}

\ifx\pdfoutput\relax\let\pdfoutput=\undefined\fi
\newcount\msipdfoutput
\ifx\pdfoutput\undefined\else
\ifcase\pdfoutput\else
\ifx\paperwidth\undefined\else
\ifdim\paperheight=0pt\relax\else\pdfpageheight\paperheight\fi
\ifdim\paperwidth=0pt\relax\else\pdfpagewidth\paperwidth\fi
\fi\fi\fi
\begin{document}
\title{Topologically non-trivial configurations in the 4d Einstein--nonlinear
$\sigma$-model system}
\author{Fabrizio Canfora}
\email{canfora@cecs.cl}
\affiliation{Centro de Estudios Cient\'{\i}ficos (CECS), Casilla 1469, Valdivia, Chile}
\author{Nikolaos Dimakis}
\email{nsdimakis@gmail.com}
\affiliation{Instituto de Ciencias F\'{\i}sicas y Matem\'{a}ticas, Universidad Austral de
Chile, Valdivia, Chile}
\author{Andronikos Paliathanasis}
\email{anpaliat@phys.uoa.gr}
\affiliation{Instituto de Ciencias F\'{\i}sicas y Matem\'{a}ticas, Universidad Austral de
Chile, Valdivia, Chile}
\affiliation{Institute of Systems Science, Durban University of Technology, PO Box 1334,
Durban 4000, Republic of South Africa}

\begin{abstract}
We construct exact, regular and topologically non-trivial\ configurations of
the coupled Einstein-nonlinear sigma model in (3+1) dimensions. The ansatz for
the nonlinear $SU(2)$ field is regular everywhere and circumvents Derrick's
theorem because it depends explicitly on time, but in such a way that its
energy-momentum tensor is compatible with a stationary metric. Moreover, the
$SU(2)$ configuration cannot be continuously deformed to the trivial Pion
vacuum as it possesses a non-trivial winding number. We reduce the full
coupled 4D Einstein nonlinear sigma model system to a single second order
ordinary differential equation. When the cosmological constant vanishes, such
master equation can be further reduced to an Abel equation. Two interesting
regular solutions correspond to a stationary traversable wormhole (whose only
\textquotedblleft exotic matter" is a negative cosmological constant) and a
(3+1)-dimensional cylinder whose (2+1)-dimensional section is a Lorentzian
squashed sphere. The Klein-Gordon equation in these two families of spacetimes
can be solved in terms of special functions. The angular equation gives rise
to the Jacobi polynomials while the radial equation belongs to the
Poschl-Teller family. The solvability of the Poschl-Teller problem implies
non-trivial quantization conditions on the parameters of the theory.

\end{abstract}
\maketitle
\date{\today}

\section{Introduction}

The nonlinear sigma model is an important effective field theory with many
applications ranging from quantum field theory to statistical mechanics
systems like quantum magnetism, the quantum hall effect, meson interactions,
super fluid $^{3}$He, and string theory \cite{manton}. The most relevant
application for the $SU(2)$ nonlinear sigma model in particle physics is the
description of the low-energy dynamics of Pions in 3+1 dimensions (see for
instance \cite{example}; for a detailed review \cite{nair}).

The nonlinear sigma models do not admit static globally regular soliton
solutions with non-trivial topological properties in flat, topologically
trivial (3+1)-dimensional space-times. This can be shown using the Derrick's
scaling argument \cite{derrick}. There are two useful strategies to avoid
Derrick's argument in the nonlinear sigma model: the first is to search for a
time-periodic ansatz such that the energy-density of the configuration is
still static, as it happens for Boson stars \cite{bosonstar} in the simpler
case of a $U(1)$ charged scalar field (for a detailed review see
\cite{bosonstar2}). The second idea is to couple the nonlinear sigma model
with the Einstein theory. In the present paper, we will exploit both of them.
The possibility of applying the first idea is prompted by the recent
generalization of the boson star ansatz to $SU(2)$-valued scalar fields
\cite{canfora,canfora2,canfora3,canfora4}, \cite{canfora5}, \cite{canfora7},
\cite{canfora8} and \cite{canfora9}. On the other hand, the idea of coupling
the nonlinear sigma model to Einstein gravity has been tried, but mostly
relying on numerical analyses because of the complexity of the system.
Numerical solutions for the Einstein-nonlinear $SU\left(  2\right)  $ $\sigma
$-model with cosmological constant were derived in \cite{solitonscosmo}.
Recently exact solutions were presented for the time-dependent gravitating
Einstein-Skyrme model \cite{cosmskyrme} by using the ansatz of \cite{canfora}.

In the present paper, we construct two families of analytic topologically
non-trivial\footnote{Topologically non-trivial in the sense that the ansatz
for the $SU(2)$-valued matter field possesses a non-vanishing third homotopy
class. This implies that it cannot be deformed continuously to the trivial
Pions vacuum.} solutions of the Einstein-nonlinear sigma model system (both
with and without $\Lambda$) with quite novel geometrical properties.

In the case in which $\Lambda<0$, some of these configurations (shortly
described in \cite{ACZ}) are smooth and regular everywhere describing
wormholes with non-vanishing NUT parameter \cite{NUTcharge1, NUTcharge2,
NUTcharge5} (for a detailed discussion of the NUT geometry, see
\cite{NUTcharge3, NUTcharge4, griffpod} and references therein).

It is worth to emphasize that in \cite{clement} \cite{clement1} it has been
shown that some typical obstructions to accept solutions with NUT parameter as
physically relevant can be removed.

Thus, the present wormholes are supported by a cloud of interacting Pions and
the only exotic ingredient needed for the construction is a negative
cosmological constant (which can hardly be considered as an exotic ingredient).

Moreover, in the $\Lambda=0$ case, we also construct a second family of
regular configurations corresponding to a (3+1) dimensional cylinder whose
(2+1)-dimensional sections are Lorentzian squashed sphere. Also in this case
the matter field supporting the configuration cannot be deformed continuously
to the trivial Pions vacuum.

We exploit the gauge invariance of gravitational systems in order to reduce
the full coupled system of Einstein nonlinear sigma model equations to a
single second order ordinary differential equation (henceforth ODE) which, in
the vanishing $\Lambda$ case, can be further reduced to a first order Abel
equation. Due to the complexity of the 3+1 dimensional Einstein $SU(2)$ sigma
model, the reduction of the full coupled system to a single second order ODE
is a very useful property that can be used by other researchers working in the
field. A further non-trivial characteristic of these configurations is that
the Klein-Gordon equation on these backgrounds can be fully integrated in
terms of special functions.

This paper is organized as follows: in the second section, the action of the
system is introduced while in the third, the ansatz for the $SU(2)$ valued
matter field is described and the field equations are extracted. In the fourth
section, we derive out of the field equations of motion a single second order
master equation and interesting particular solutions are analyzed. In section
five, the intriguing properties of the Klein-Gordon equation on the two
families of regular configurations are disclosed. In the last section, some
conclusions are drawn.

\section{The Action}

We consider the Einstein theory minimally coupled with the $SU(2)$-valued
nonlinear sigma model system in four dimensions. The nonlinear sigma model
describes the low-energy dynamics of pions, whose degrees of freedom are
encoded in an SU(2) group-valued scalar field $U$ \cite{nair}. The action of
the system is (we will follow the notation of \cite{ACZ})
\begin{equation}
S=S_{\mathrm{G}}+S_{\mathrm{Pions}}, \label{action0}%
\end{equation}
where the gravitational action $S_{\mathrm{G}}$ and the nonlinear sigma model
action $S_{\mathrm{Pions}}$ are given by
\begin{align}
S_{\mathrm{G}}=  &  \frac{1}{16\pi G}\int d^{4}x\sqrt{-g}(\mathcal{R}%
-2\Lambda),\label{sky1}\\
S_{\mathrm{Pions}}=  &  \frac{K}{2}\int d^{4}x\sqrt{-g}\mathrm{Tr}\left(
R^{\mu}R_{\mu}\right)  \ ,R_{\mu}=U^{-1}\nabla_{\mu}U\ ,\label{sky2}\\
U  &  \in SU(2)\ ,\ \ R_{\mu}=R_{\mu}^{j}t_{j}\ ,\ \ t_{j}=i\sigma_{j}\ ,
\label{sky2.5}%
\end{align}
where $\mathcal{R}$ is the Ricci scalar, $G$ is Newton's constant, the
parameter $K$ $\left(  >0\right)  $ is experimentally fixed and $\sigma_{j}$
are the Pauli matrices. In our conventions $c=\hbar=1$, the space-time
signature is $(-,+,+,+)$ (although we will shortly discuss Euclidean
solutions) and Greek indices run over space-time. Here $\Lambda$ is the
cosmological constant.

As it is well known the non-linear sigma model can be seen as the $\lambda=0$
limit ($\lambda$ being the Skyrme coupling constant) of the Skyrme model
(which, on flat spaces, allows the existence of non-trivial topological
configurations). We have decided to consider here only the non-linear sigma
model since our analysis provides with explicit examples of how the coupling
with Einstein theory allows the non-linear sigma model to have smooth
configurations with non-vanishing topological charge.

The resulting Einstein equations which follow from the previous action are
\begin{equation}
G_{\mu\nu}+\Lambda g_{\mu\nu}=\kappa T_{\mu\nu}, \label{einstein}%
\end{equation}
where $G_{\mu\nu}$ is the Einstein tensor and $\kappa$ the gravitational
constant. The stress-energy tensor is
\begin{equation}
T_{\mu\nu}=-\frac{K}{2}\mathrm{Tr}\left(  R_{\mu}R_{\nu}-\frac{1}{2}g_{\mu\nu
}R^{\alpha}R_{\alpha}\right)  \,,\nonumber
\end{equation}
which, for the nonlinear sigma model can be seen to satisfy the dominant and
strong energy conditions \cite{gibbons2003}. Finally, the matter field
equations are
\begin{equation}
\nabla^{\mu}R_{\mu}=0. \label{nonlinearsigma1}%
\end{equation}

We adopt the standard parametrization of the $SU(2)$-valued scalar $U(x^{\mu
})$
\begin{equation}
U^{\pm1}(x^{\mu})=Y^{0}(x^{\mu})\mathbb{\mathbf{I}}\pm Y^{i}(x^{\mu}%
)t_{i}\ ,\ \ \left(  Y^{0}\right)  ^{2}+Y^{i}Y_{i}=1\,, \label{standnorm}%
\end{equation}
where $\mathbb{\mathbf{I}}$ is the $2\times2$ identity. The last equality
implies that $Y^{A}:=(Y^{0},Y^{i})$ is a unit vector in a three sphere, which
is naturally accounted for by writing
\begin{align}
Y^{0}  &  =\cos\alpha\ ,\ Y^{i}=n^{i}\cdot\sin\alpha\ ,\label{pions1}\\
n^{1}  &  =\sin\Theta\cos\Phi\ ,\ \ n^{2}=\sin\Theta\sin\Phi\ ,n^{3}%
=\cos\Theta\ . \label{pions2}%
\end{align}

\section{The field equations}

The full coupled Einstein-nonlinear sigma model system of equations
(\ref{einstein}, \ref{nonlinearsigma1}) can be consistently analyzed on the
following family of metrics
\begin{equation}
ds^{2}=-F(r)\left(  dt+\cos\theta d\varphi\right)  ^{2}+ N(r)^{2} dr^{2}%
+\rho^{2}(r)\left(  d\theta^{2}+\sin^{2}\theta d\varphi^{2}\right)  \ \,.
\label{metricnew}%
\end{equation}
As already on flat space-times the field equations of the nonlinear sigma
model are very complicated, it is necessary to take great care in choosing the
ansatz for the $SU(2)$-valued scalar field $U$.

Following the recipe of \cite{canfora,canfora2,canfora3,canfora4},
\cite{canfora5}, \cite{canfora7}, \cite{canfora8} and \cite{canfora9}, in the
reference \cite{ACZ} the following ansatz time-dependent for $\alpha$,
$\Theta$\ and $\Phi$\ has been proposed:
\begin{equation}
\Phi=\frac{t+\varphi}{2}\ ,\ \ \tan\Theta=\frac{\cot\left(  \frac{\theta}%
{2}\right)  }{\cos\left(  \frac{t-\varphi}{2}\right)  }\ ,\ \ \tan\alpha
=\frac{\sqrt{1+\tan^{2}\Theta}}{\tan\left(  \frac{t-\varphi}{2}\right)  }\ .
\label{pions2.25}%
\end{equation}
One can verify directly that\textit{\ in any metric of the form in Eq}.
(\ref{metricnew}), a Pionic configuration of the form in Eqs. (\ref{pions2.25}%
), (\ref{pions1}) and (\ref{pions2})\textit{\ identically satisfies the
nonlinear sigma model field equations} Eq. (\ref{nonlinearsigma1})\textit{.}
It is also worth to emphasize that the present ansatz is topologically
non-trivial as it has a non-trivial winding number along the $\left\{
r=const\right\}  $ surfaces:%
\[
W=-\frac{1}{24\pi^{2}}\int_{\left\{  r=const\right\}  }tr\left[  \left(
U^{-1}dU\right)  ^{3}\right]  =-\frac{1}{2\pi^{2}}\int_{\left\{
r=const\right\}  }\sin^{2}\alpha\sin\Theta d\alpha d\Theta d\Phi\neq0\ .
\]
Therefore, the present configuration cannot be deformed continuously to the
trivial Pions vacuum $U_{0}=\mathbf{1}$.

Furthermore, it is worth to emphasize that the non-vanishing winding number of
the present configurations is closely related to the non-standard asymptotic
behavior which characterizes both types of solutions constructed in the next
sections. Indeed, it is easy to check that if the asymptotic behavior of the
metric would be AdS, dS or Minkowski then the present configurations would
have vanishing winding number. Such configuration belongs to a different
sector with respect to the usual ones analyzed in the literature on the
Einstein non-linear sigma model system\footnote{As in most of the examples
analyzed in the literature the Pions have vanishing winding number.}.

Despite the fact that such configurations explicitly depend on the time-like
coordinate, the energy momentum tensor is compatible with the above stationary
metric:
\begin{align}
T_{t}^{t} &  =-\frac{K(2F+\rho^{2})}{8F\rho^{2}},\,T_{r}^{r}=-\frac
{K(2F-\rho^{2})}{8F\rho^{2}},\nonumber\\
T_{\theta}^{\theta} &  =T_{\varphi}^{\varphi}=\frac{K}{8F}\ ,\ T_{t}^{\varphi
}=-\frac{K\left(  F+\rho^{2}\right)  }{4F\rho^{2}}\cos\theta
\ .\label{tmunu1.1}%
\end{align}
Hence, the ansatz in Eqs. (\ref{pions2.25}), (\ref{pions1}) and (\ref{pions2})
avoids the Derrick theorem since it is explicitly time-dependent and, at the
same time, compatible with a stationary space-time. Thus, the present ansatz
is the $SU(2)$ generalization of the well-known bosons-stars like ansatz in
which a $U(1)$ scalar field depends explicitly on time in such a way that the
energy-momentum tensor does not. However, the present case is worth to be
further analyzed for, at least, two reasons: first of all, the matter field
corresponds to Pions and secondly, $SU(2)$-valued matter fields may possess
non-trivial topological properties. It is also worth to emphasize that the
above $T_{\mu\nu}$ has positive energy density. Indeed, as is well known
\cite{gibbons2003}, the $T_{\mu}^{\nu}$ of the nonlinear sigma model satisfies
both the null and the weak energy conditions (as can be checked directly in
Eq. (\ref{tmunu1.1})).

A direct computation reveals that the full Einstein-nonlinear sigma model
system reduces, in this sector, to the following three equations (two second
order equations and one constraint):%
\begin{align}
0 =  &  16\rho^{3}F N \rho^{\prime\prime} + 8\rho^{2}F N (\rho^{\prime})^{2}-
16 \rho^{3} F N^{\prime} \rho^{\prime} + N^{3} \left[  K\kappa\rho^{2}\left(
2F+\rho^{2}\right)  +2F\left(  4\rho^{2}\left(  \Lambda\rho^{2}-1\right)
-3F\right)  \right] \label{prop1}\\
0 =  &  -8\rho^{2}F\, (\rho^{\prime})^{2}-8\rho^{3} \rho^{\prime}F^{\prime}+
N^{2} \left[  K\kappa\rho^{2}\left(  \rho^{2}-2F\right)  -2F\left(  4\rho
^{2}\left(  \Lambda\rho^{2}-1\right)  -F\right)  \right] \label{prop2}\\
0=  &  -8\rho^{3}F^{2} N \rho^{\prime\prime} -4\rho^{4}F N F^{\prime\prime}
-4\rho^{3}F N \rho^{\prime}F^{\prime}+ 2\rho^{4}N \, ( F^{\prime})^{2} +
4\rho^{3} F\left(  2F N^{\prime}\rho^{\prime}+\rho N^{\prime} F^{\prime
}\right) \nonumber\\
&  +F N^{3} \left(  K\kappa\rho^{4}-2F\left(  4\Lambda\rho^{4}+F\right)
\right)  \label{prop3}%
\end{align}
where Eq. (\ref{prop1}),(\ref{prop2}) and (\ref{prop3}) correspond to the
$t-t$, $r-r$ and $\theta-\theta$ components of the Einstein equations
respectively, while the prime denotes differentiation with respect to $r$.

It can be easily seen that these remaining Einstein equations are not
independent from one another. Indeed, a direct computation shows that the
total derivative of \eqref{prop2} is a combination of \eqref{prop1} and
\eqref{prop3}. {Moreover it is straightforward to see that the field
equations can be derived from the variation of a Lagrangian
}$\mathcal{L}=L\left(  N,F,F^{\prime},\rho,\rho^{\prime}\right)  $. {Due the reparametrization invariance of this Lagrangian, equations (\ref{prop1})-(\ref{prop3}) form a singular dynamical
system.} From the theory of constrained systems we know that equation
\eqref{prop2} - since it corresponds to a first class constraint in the
Hamiltonian formalism of the theory - removes a full degree of freedom from
the system (classic references are \cite{Dirac} and \cite{Sund}). Thus,
between $F$ and $\rho$ only one can be considered as a true physical degree of
freedom and its evolution can be described by a single second order equation,
which we derive in the following section.

{In order to see the classical analogue of that through a gauge fixing process, consider without loss
of generality that }$N=N\left(  F,\rho\right)  ${, then the field
equations describe a classical system of two degrees of freedom in which
(\ref{prop2}) is the Hamiltonian function expressed in velocity phase space variables. It can be seen a conservation law with a fixed value, which means that it constrains the solution. Indeed, if we solve the two second-order differential equations, one of the four
integration constants is constraint from (\ref{prop2}). Moreover, because
(\ref{prop2}) is the Hamiltonian function it means that the dynamical system
is autonomous and the solution is invariant under translations of the
independent variable, in our consideration the radius. Hence, a second
integration constant can be eliminated, which means that the there are only
two integration constants and consequently only one free degree of freedom,
which means that there exists a second-order differential equation which
describes equivalently the solution of the field equations.}

\section{Derivation of the master equation}

Let us avoid the typical fixing condition $N=1$ and exploit the fact that the
constraint equation \eqref{prop2} is algebraic with respect to $N$; its
solution is:
\begin{equation}
N(r)=\pm2\rho\left(  \frac{\rho^{\prime}\left(  \rho F\right)  ^{\prime}%
}{F^{2}+\frac{1}{2}\bar{K}\rho^{4}-F\rho^{2}\left(  \bar{K}+4\Lambda\rho
^{2}-4\right)  }\right)  ^{\frac{1}{2}},\label{lapse}%
\end{equation}
where from now on we consider for simplicity a single constant $\bar
{K}=K\kappa$. Upon substitution of \eqref{lapse} in \eqref{prop1} and
\eqref{prop3}, they both reduce to a single equation\footnote{This is a
further confirmation of the fact that among $F$ and $\rho$ only one can be
considered as a true physical degree of freedom.} for of $\rho$ and $F$,
namely
\begin{equation}%
\begin{split}
&  F\rho^{2}F^{\prime}\left(  2F^{2}+\bar{K}\rho^{4}-2F\rho^{2}\left(  \bar
{K}+4\Lambda\rho^{2}-4\right)  \right)  \rho^{\prime\prime}-F\rho^{2}%
\rho^{\prime}\left(  2F^{2}+\bar{K}\rho^{4}-2F\rho^{2}\left(  \bar{K}%
+4\Lambda\rho^{2}-4\right)  \right)  F^{\prime\prime}\\
&  +\rho^{\prime}\Big[F\rho F^{\prime}\rho^{\prime}\left(  4(\bar{K}%
-4)F\rho^{2}-14F^{2}+3\bar{K}\rho^{4}\right)  +\rho^{2}(F^{\prime})^{2}\left(
\bar{K}\rho^{4}-2F^{2}\right)  \\
&  +4F^{2}(\rho^{\prime})^{2}\left(  \rho^{4}(\bar{K}-4\Lambda F)-2F^{2}%
\right)  \Big]=0.
\end{split}
\label{master0}%
\end{equation}
We have to note here that solving \eqref{prop2} with respect to $N(r)$ does
not constitute a fixing choice, since the constraint equation must be
satisfied in any lapse. As a result, we still have the freedom to gauge fix
the system by choosing either $\rho$ or $F$ to be some explicit function of
$r$. We make the following choice regarding $\rho$ together with a
reparametrization condition for $F$, by introducing a new function $g(r)$ in
its place\footnote{The reasoning behind this choice and parametrization can be
traced to a mini-superspace analysis of the system and the fact that it
simplifies a non-local integral of motion constructed out of a conformal
vector of the mini-superspace metric. For similar technics used in a
mini-superspace context in order to achieve integrability see for example
\cite{tchris1}, \cite{dim2} and in regular systems \cite{mgrg}.}:
\begin{equation}
\rho(r)=r,\quad F=\frac{g(r)}{r}.\label{simpcond}%
\end{equation}
Then, equation \eqref{master0} simplifies to the following second order
differential equation
\begin{equation}
rg\left(  \bar{K}r^{6}-2r^{3}\left(  \bar{K}-4+4r^{2}\Lambda\right)
g+2g^{2}\right)  g^{\prime\prime}-g^{\prime}\left(  3\bar{K}r^{6}g-6g^{3}%
+\bar{K}r^{7}g^{\prime}-2rg^{2}\left(  8r^{4}\Lambda+g^{\prime}\right)
\right)  =0.\label{finsn4}%
\end{equation}
As a result, we have succeeded in reducing the initial system of equations
into a problem of finding the solution of a single non-autonomous second order
ODE. As it is shown in the examples below, \textit{all the metric components
can be constructed once the solution of the above master equation is known}.

In particular, once a solution of the above master equation (\ref{finsn4}) is
known, one can construct a solution of the full system made by Eqs.
(\ref{prop1}), (\ref{prop2}) and (\ref{prop3}) just using Eqs. (\ref{simpcond}%
) and (\ref{lapse}).

What is more, equation \eqref{finsn4} can be further reduced (when $\Lambda
=0$) to a first order ODE of Abel type.

In the remaining of this section we shall study sets of particular solution of
\eqref{finsn4} as well as its group invariant transformations and specifically
the Lie point symmetries.

It is worth to emphasize that the reduction of the full coupled 4D Einstein
non-linear sigma model system in a topologically non-trivial sector to a
single second order ODE is quite an achievement in itself and will be very
useful for researchers in the field. For instance, the above master equation
is a very good starting point to analyze whether or not there are black holes
in this sector of the theory. We hope to come back on this issue in a future publication.

\subsection{Solution from Lie symmetry}

Equation (\ref{finsn4}) is of the general form $g^{\prime\prime}=\Omega\left(
r,g,g^{\prime}\right)  $. It is defined in the jet space $A_{T}=\left\{
r,g,g^{\prime},g^{\prime\prime}\right\}  ,$ hence in order to be invariant
under an infinitesimal point transformation of the form%
\begin{equation}
\bar{r}=r+\varepsilon\xi\left(  r,g\right)  ~,~\bar{g}=g+\varepsilon
\eta\left(  r,g\right)  , \label{sym.01}%
\end{equation}
whose generator is $X= \xi\frac{\partial}{\partial r} + \eta\frac{\partial
}{\partial g}$, the following condition should hold%
\begin{equation}
X^{\left[  2\right]  }\left(  g^{\prime\prime}-\Omega\right)  =0, \quad
\quad\mathrm{mod} \quad g^{\prime\prime}= \Omega, \label{sym.02}%
\end{equation}
where $X^{[2]}$ is the second prolongation of $X$, i.e. the extension of the
generator into the jet space $A_{T}$ (for the rigorous mathematical
definitions, as well as the general theory of Lie point symmetries, we refer
the interested reader to various textbooks on the subject e.g. \cite{Bluman},
\cite{Olver}). If \eqref{sym.02} is satisfied, then $X$ generates a Lie point
symmetry for the differential equation. Its existence provides a point
transformation under which equation (\ref{finsn4}) can be either transformed
to an autonomous or reduced to a first order equation.

From the symmetry condition (\ref{sym.02}) we find that equation
(\ref{finsn4}) admits a Lie point symmetry if an only if the cosmological
constant is zero. The corresponding symmetry vector is found to be
\begin{equation}
X=\partial_{r}+3g\partial_{g}%
\end{equation}
and can be used to further simplify the equation when $\Lambda=0$.

By using this symmetry vector we can introduce the normal coordinates,
\begin{equation}
r=\exp\left(  \frac{R}{3}\right)  ,~\text{and,~}g=e^{R}S\left(  R\right)  ,
\label{sym.02b}%
\end{equation}
with the help of which equation (\ref{finsn4}) assumes the following form%
\begin{equation}
3S_{,RR}+\left(  2S+5S_{,R}\right)  -\frac{3\left(  \bar{K}-2S^{2}\right)
\left(  S+S_{,R}\right)  \left(  2S+S_{,R}\right)  }{S\left(  \bar{K}-2\left(
\bar{K}-4\right)  S+2S^{2}\right)  }=0. \label{sym.03}%
\end{equation}
The latter is an autonomous second order differential equation due to the fact
that the variables we introduced transformed the symmetry vector into
$X=\partial_{R}$.

\subsubsection{The Abel master equation with $\Lambda=0$}

In the sector with vanishing cosmological constant it is possible to further
simplify the master equation in Eq. (\ref{finsn4}). Indeed, from the
application of the differential invariants, \eqref{sym.03} can be reduced to a
first order Abel equation of the second kind
\begin{equation}
v_{,z}=\lambda_{0}\left(  z\right)  \left[  \lambda_{1}\left(  z\right)
v+\frac{\lambda_{2}\left(  z\right)  }{v}+\lambda_{3}\left(  z\right)
\right]  , \label{sym.03a}%
\end{equation}
where $z=S\left(  R\right)  $, $v\left(  z\right)  =S_{,R}$ or $R=\int\frac
{1}{v\left(  z\right)  }dz+R_{0}$. Coefficients $\lambda_{0,1,2,3}$ are as
follows%
\begin{equation}
\lambda_{0}\left(  z\right)  =\left[  3z\left(  2\left(  \bar{K}-4\right)
z-2z^{2}-\bar{K}\right)  \right]  ^{-1}\text{,~}\lambda_{1}\left(  z\right)
=\left(  6z^{2}-3\bar{K}\right)  , \label{sym.03b}%
\end{equation}%
\begin{equation}
\lambda_{2}\left(  z\right)  =4\left(  4z^{4}+\left(  4-\bar{K}\right)
z^{3}-\bar{K}z^{2}\right)  ,
\end{equation}%
\begin{equation}
\lambda_{3}\left(  z\right)  =28z^{3}+10\left(  4-\bar{K}\right)  z^{2}%
-4\bar{K}z. \label{sym.03d}%
\end{equation}

Of course with a change of variables equation (\ref{sym.03a}) can be written
as an Abel equation of the first kind \cite{Zai2}.

In the following we study some particular solutions.

\subsection{Particular solutions}

It can be easily verified that equation \eqref{finsn4} admits a power law
solution that is given by
\begin{equation}
g(r)=\sigma\,r^{3}%
\end{equation}
whenever $\sigma=\frac{\bar{K}}{4}$ or $\sigma=-1$. We begin our analysis by
examining the former case. Solutions of that form indicate that $F=\sigma
\rho^{2}$ as we can see from \eqref{simpcond}. In what follows we study those
solutions in the original coordinates of the space-time.

\subsubsection{The $\sigma=\frac{\bar{K}}{4}$ case: the wormhole}

This solution corresponds to the Lorentzian wormhole constructed in
\cite{ACZ}. Here, as we can observe from \eqref{lapse} and \eqref{simpcond},
it is expressed in a gauge where the \textquotedblleft lapse" $N(r)$ of the
metric is
\begin{equation}
N(r) = \pm\frac{4\sqrt{3}}{\sqrt{24-3\bar{K}-16\Lambda r^{2}}}.
\end{equation}
At this point, to see how the solution is expressed in the original variables,
we can perform a transformation $r\rightarrow\tilde{r}$ that returns us to the
gauge $N(\tilde{r})=1$. For the case when $\Lambda\neq0$ and $\bar{K}\neq8$
this is achieved by
\begin{equation}
\int\!\!N(r)dr=\tilde{r}\Rightarrow r=\lambda\sqrt{\frac{3(\bar{K}%
-8)}{16(-\Lambda)}}\cosh\left(  \frac{\sqrt{-\Lambda}\,\tilde{r}}{\sqrt{3}%
}\right)  +\sqrt{1-\lambda^{2}}\sqrt{\frac{3(\bar{K}-8)}{16\Lambda}}%
\sinh\left(  \frac{\sqrt{-\Lambda}\,\tilde{r}}{\sqrt{3}}\right)
\label{transfg1}%
\end{equation}
where $\lambda$ is an integration constant which can be either real or
imaginary, it depends on the behavior of the rest of the parameters $\bar{K}$
and $\Lambda$. We have to note here that the original configuration space
variables $F$ and $\rho$ that appear on the metric are now given as
\begin{subequations}
\begin{align}
\rho(\tilde{r})  &  =r(\tilde{r})=\lambda\sqrt{\frac{3(\bar{K}-8)}%
{16(-\Lambda)}}\cosh\left(  \frac{\sqrt{-\Lambda}\,\tilde{r}}{\sqrt{3}%
}\right)  +\sqrt{1-\lambda^{2}}\sqrt{\frac{3(\bar{K}-8)}{16\Lambda}}%
\sinh\left(  \frac{\sqrt{-\Lambda}\,\tilde{r}}{\sqrt{3}}\right)
\label{rhosolution}\\
F(\tilde{r})  &  =\frac{g(\tilde{r})}{r(\tilde{r})}=\frac{\bar{K}}{4}%
r(\tilde{r})^{2}=\frac{\bar{K}}{4}\rho^{2}(\tilde{r})
\end{align}
Since only $\rho^{2}$ appears in the metric, all combinations for which
\eqref{transfg1} is real (or pure imaginary) are acceptable for a Lorentzian
(or Euclidean) signature solution. Let us note, that for the rest of this
subsection - and for reasons of simplicity - we shall omit the tilde,
understanding that we are going to restrict our analysis solely in the gauge
$N=1$ in which we decide to symbolize the dynamical variable from now on as
$r$.

Let us consider for example the case when $\bar{K}>8$, $\Lambda<0$ and at the
same time $|\lambda|\geq1$. By setting $|\lambda|=\cosh m$ inside solution
\eqref{rhosolution} and applying the appropriate trigonometric identity we get%

\end{subequations}
\begin{equation}
\rho(r)=\pm\frac{1}{4}\sqrt{\frac{3(\bar{K}-8)}{-\Lambda}}\cosh\left(
\sqrt{\frac{-\Lambda}{3}}\,r\pm m\right)  \label{rhoworm}%
\end{equation}
with the $\pm$ signs not being of importance, since only $\rho^{2}$ appears on
the metric and $m$ being able to be either positive or negative. By checking
the metric components it is obvious that this latter constant is not even
essential for the geometry because it can be easily absorbed by a translation
in the $r$ variable. Hence, without any loss of generality, we can assume $m$
to be equal to zero and write the most general solution, for this specific
range of validity of the parameters $\bar{K}$ and $\Lambda$:
\begin{subequations}
\label{coshsol}%
\begin{align}
\rho(r)  &  =\frac{1}{4}\sqrt{\frac{3(\bar{K}-8)}{-\Lambda}}\cosh\left(
\sqrt{\frac{-\Lambda}{3}}\,r\right) \label{wormh1}\\
F(r)  &  =\frac{\bar{K}}{4}\rho(r)^{2}\ ,\ \ \bar{K}-8>0\ . \label{wormh2}%
\end{align}

As it has been discussed in \cite{ACZ}, the metric corresponding to Eq.
(\ref{rhoworm}) is a traversable Lorentzian wormhole constructed using only
ingredients from the standard model and a negative cosmological constant. To
analyze the global structure of the solution it is useful to compare it with
the NUT-AdS line element:
\end{subequations}
\begin{equation}
ds^{2}=-f(r)\left(  dt+2n\cos\theta d\phi\right)  ^{2}+\frac{dr^{2}}%
{f(r)}+(r^{2}+n^{2})d\Omega_{2}^{2}, \label{nutmet}%
\end{equation}
where
\begin{equation}
f(r)=\frac{r^{2}+N^{2}+\ell^{-2}\left(  r^{4}-6N^{2}r^{2}-3N^{4}\right)
}{r^{2}-N^{2}} \label{finnut}%
\end{equation}
which in the asymptotical limit $r\rightarrow+\infty$ can be written as
\begin{equation}
f_{\infty}(r)=\ell^{-2}r^{2}+(1-5\ell^{-2}N^{2})+\mathcal{O}(r^{-2}).
\end{equation}
The corresponding asymptotical form of line element \eqref{nutmet} can be
transformed under $r=\ell\left(  1-\frac{5N^{2}}{\ell^{2}}\right)  ^{1/2}%
\sinh(\bar{r}/\ell)$ (assuming $\ell^{-2}>5N^{2}$ and $n=(\ell^{2}%
-5N^{2})^{1/2}$) into
\begin{equation}
ds^{2}=-\frac{\left(  \ell^{2}-5N^{2}\right)  \cosh^{2}(\bar{r}/\ell)}{l^{2}%
}\left(  dt+2(\ell^{2}-5N^{2})^{1/2}\cos\theta d\phi\right)  ^{2}+d\bar{r}%
^{2}+\left(  \ell^{2}-5N^{2}\right)  \cosh^{2}(\bar{r}/\ell)d\Omega_{2}^{2},
\label{nutmet2}%
\end{equation}
which belongs to the general form of our solution. Thus, the asymptotic form
of a NUT-AdS solution coincides with our result. The difference being of
course that the present solution has two identical asymptotic regions for
$r\rightarrow\pm\infty$.

Moreover, the ensuing space-time we get from \eqref{metricnew} does not
exhibit any curvature singularity. For instance, the Kretschmann scalar is%
\begin{equation}
\label{coshsol1}%
\begin{split}
K_{s}=R_{\kappa\lambda\mu\nu}R^{\kappa\lambda\mu\nu}=  &  \frac{\Lambda^{2}%
}{9(\bar{K}-8)^{2}\cosh^{4}\left(  \sqrt{\frac{-\Lambda}{3}}\,r\right)
}\left[  3(\bar{K}-8)^{2}\cosh\left(  4\sqrt{\frac{-\Lambda}{3}}\,r\right)
\right. \\
&  \left.  -4(\bar{K}-8)(\bar{K}+16)\cosh\left(  2\sqrt{\frac{-\Lambda}{3}%
}\,r\right)  +57\bar{K}^{2}+272\,\bar{K}+1088\right]
\end{split}
\end{equation}
and it is regular for all $r\in\mathbb{R}$. The special properties of the
Klein-Gordon equation on this metric will be described in the next sections.

For the sake of brevity we state in Table \ref{tablee1} all the admissible
Lorentzian solutions corresponding to the various choices of parameters. The
procedure is similar and in all cases the integration constant can be absorbed
by a simple translation in the $r$ variable.%

%TCIMACRO{\TeXButton{B}{\begin{table}[tbp] \centering}}%
%BeginExpansion
\begin{table}[tbp] \centering
%EndExpansion
\caption{Admissible values of essential parameters and singularities}%
\begin{tabular}
[c]{ccc}\hline\hline
\textbf{Parameters} & $\mathbf{\rho}\left(  r\right)  $ &
\textbf{singularities}\\\hline
$\bar{K}>8$, $\Lambda<0$ & $\frac{1}{4}\sqrt{\frac{3(\bar{K}-8)}{-\Lambda}%
}\cosh\left(  \sqrt{\frac{-\Lambda}{3}}r\right)  $ & no\\
$\bar{K}<8$, $\Lambda<0$ & $\frac{1}{4}\sqrt{\frac{3(8-\bar{K})}{-\Lambda}%
}\sinh\left(  \sqrt{\frac{-\Lambda}{3}}r\right)  $ & $r=0$\\
$\bar{K}<8$, $\Lambda>0$ & $\frac{1}{4}\sqrt{\frac{3(8-\bar{K})}{\Lambda}}%
\sin\left(  \sqrt{\frac{\Lambda}{3}}r\right)  $ & $r=\sqrt{\frac{3}{\Lambda}%
}k\pi,k\in\mathbb{Z}$\\\hline\hline
\end{tabular}
\label{tablee1}%
%TCIMACRO{\TeXButton{E}{\end{table}}}%
%BeginExpansion
\end{table}%
%EndExpansion

The corresponding Euclidean solutions can be obtained in each case by simply
violating the conditions given for $\bar{K}$ in the first column of Table
\ \ref{tablee1}. For example, if we consider $\bar{K}<8$ and $\Lambda<0$ the
first solution is written as $\rho(r)=\frac{\mathbbmtt{i}}{4}\sqrt
{\frac{3(8-\bar{K})}{-\Lambda}}\cosh\left(  \sqrt{\frac{-\Lambda}{3}}r\right)
$ and leads to a Euclidean space without a singularity. An interesting pattern
may be observed in what regards these solutions: The curvature singularities
appear for $\bar{K}<8$ in the Lorentzian manifolds, while the situation is
inverted for Euclidean space-times, where they appear only for $\bar{K}>8$.

Of course we have still to examine three special cases spanned by $\Lambda=0 $
and/or $\bar{K}=8$, which were excluded in transformation \eqref{transfg1}.
Let us begin by considering $\Lambda=0$. In this case, one only needs to
perform a scaling in the radial variable to express the result in the gauge
$N=1$. The corresponding solution reads
\begin{subequations}
\begin{align}
\rho(r)  &  =\frac{1}{4}\sqrt{8-\bar{K}}\,r\\
F(r)  &  =\frac{\bar{K}}{4}\rho(r)^{2}.
\end{align}
where $\bar{K}\neq8$. The line element is Lorentzian if $\bar{K}<8$ and
Euclidean when $\bar{K}>8$. In both cases the manifold exhibits a curvature
singularity at the origin $r=0$.

When $\bar{K}=8$ (but $\Lambda\neq0$), we are led to the solution (again
expressed in the $N=1$ gauge)
\end{subequations}
\begin{subequations}
\begin{align}
\rho(r)  &  =\omega\,e^{\frac{\sqrt{-\Lambda}r}{\sqrt{3}}}\label{rnewreg}\\
F(r)  &  =2\,\rho(r)^{2}.
\end{align}
The cosmological constant has to be negative, while the constant $\omega$ can
be seen that it is not essential for the geometry and can be absorbed by a
translation in $r$. Thus, without loss of generality we can consider
$\omega=1$ (or $\omega=\mathbbmtt{i}$) for a Lorentzian (or Euclidean)
manifold. The emerging space-time is regular for $r>0$, since the Kretschmann
scalar is
\end{subequations}
\begin{equation}
K_{s}=27\,e^{-\frac{4\sqrt{-\Lambda}r}{\sqrt{3}}}+4\,\Lambda\,e^{-\frac
{2\sqrt{-\Lambda}r}{\sqrt{3}}}+\frac{8\,\Lambda^{2}}{3}.
\end{equation}
Note here that apart from \eqref{rnewreg}, the solution $\rho(r)=\omega
\,e^{-\frac{\sqrt{-\Lambda}r}{\sqrt{3}}}$ is also valid. The situation however
is inverted and the space is now regular for $r<0$.

\subsubsection{Cylindrical solution}

Another interesting solution is retrieved when both $\Lambda=0$ and $\bar
{K}=8$, which has to be examined separately. The functions $F(r)$ and
$\rho(r)$ in this case are constants, $F(r)=\frac{\bar{K}}{4}\rho^{2}%
(r)=\frac{\bar{K}}{4}\rho_{0}^{2}=$const. This leads to a regular manifold
described by the metric
\begin{equation}
ds^{2}=\rho_{0}^{2}[-2(dt+\cos\theta d\varphi)^{2}+(d\theta^{2}+\sin^{2}\theta
d\varphi^{2})]+dr^{2}, \label{metconst}%
\end{equation}
which admits five Killing fields: the four of the general line element
\eqref{metricnew}
\begin{equation}%
\begin{split}
&  \xi_{1}=\partial_{t},\quad\xi_{2}=\partial_{\phi}\\
&  \xi_{3}=\frac{\sin\phi}{\sin\theta}\partial_{t}+\cos\phi\,\partial_{\theta
}-\cot\theta\sin\phi\,\partial_{\phi},\quad\xi_{4}=-\frac{\cos\phi}{\sin
\theta}\partial_{t}+\sin\,\phi\partial_{\theta}+\cot\theta\cos\phi
\,\partial_{\phi}%
\end{split}
\label{Kilgen}%
\end{equation}
plus the obvious from the form of the metric \eqref{metconst} $\xi
_{5}=\partial_{r}$. The special properties of the Klein-Gordon equation on
this metric will be described in the next sections.

\subsubsection{The $\sigma=-1$ case}

The treatment is exactly the same as in the previous case. For brevity we
shall only demonstrate the final resulting solutions in the gauge $N=1$. In
the line element we set $F(r)=-\rho(r)^{2}$, with $\rho(r)$ being given for
various values of the parameters in table \ref{tablee2}.%

%TCIMACRO{\TeXButton{B}{\begin{table}[tbp] \centering}}%
%BeginExpansion
\begin{table}[tbp] \centering
%EndExpansion
\caption{Admissible values of essential parameters and singularities}%
\begin{tabular}
[c]{ccc}\hline\hline
\textbf{Parameters} & $\mathbf{\rho(r)}$ & \textbf{singularities}\\\hline
$\bar{K}>2$, $\Lambda<0$ & $\frac{1}{2}\sqrt{\frac{3(\bar{K}-2)}{-2\Lambda}%
}\cosh\left(  \sqrt{\frac{-\Lambda}{3}}r\right)  $ & no\\
$\bar{K}<2$, $\Lambda<0$ & $\frac{1}{2}\sqrt{\frac{3(2-\bar{K})}{-2\Lambda}%
}\sinh\left(  \sqrt{\frac{-\Lambda}{3}}r\right)  $ & $r=0$\\
$\bar{K}<2$, $\Lambda>0$ & $\frac{1}{2}\sqrt{\frac{3(2-\bar{K})}{2\Lambda}%
}\sin\left(  \sqrt{\frac{\Lambda}{3}}r\right)  $ & $r=\sqrt{\frac{3}{\Lambda}%
}k\pi,k\in\mathbb{Z}$\\\hline\hline
\end{tabular}
\label{tablee2}%
%TCIMACRO{\TeXButton{E}{\end{table}}}%
%BeginExpansion
\end{table}%
%EndExpansion

All the above solutions lead to Euclidean line elements. Lorentzian solutions
can be obtained by letting $\bar{K}$ possess values that violate the
conditions of the first column. However, the signature of the metric becomes
$(-,+,-,-)$, indicating that $r$ has effectively assumed the role of the time
parameter, hence this solutions should rather be considered as cosmological
and we refrain from their analysis here.

From the geometrical perspective the solutions expressed in the tabular are
not different from the corresponding Euclidean solutions of the previous
sections, since only a combination of constants differs in the line element.
Nevertheless, it is to be noted that the critical value for the constant
$\bar{K}$ in the line element has changed from eight to two.

Finally, we conclude this analysis by presenting the particular solutions for
three special cases that arise again: $\Lambda=0$ and/or $\bar{K}=2$. Let us
start with the cosmological constant being zero while $\bar{K}\neq2$. Then,
the function $\rho(r)$ reads
\begin{equation}
\rho(r)=\frac{1}{2}\sqrt{\frac{2-\bar{K}}{2}}\,r
\end{equation}
and the resulting manifold is Euclidean when $\bar{K}<2$ (the Lorentzian case
$\bar{K}>2$ being again cosmological) with a curvature singularity at the
origin $r=0$.

The situation when $\Lambda\neq0$ but $\bar{K}=2$ is quite different. The
solution is being given by
\begin{equation}
\rho(r) = \sigma\, e^{\frac{\sqrt{-\Lambda} r}{\sqrt{3}}}%
\end{equation}
with $\sigma$ being an integration constant which if it is real the resulting
line element is Euclidean and it is given by
\begin{equation}
ds^{2} = e^{\frac{2 \sqrt{-\Lambda} r}{\sqrt{3}}} dt^{2} + e^{\frac{2
\sqrt{-\Lambda} r}{\sqrt{3}}}\cos\theta\, dt \, d\phi+ dr^{2} +e^{\frac{2
\sqrt{-\Lambda} r}{\sqrt{3}}} (d\theta^{2} +d\phi^{2}).
\end{equation}
The constant $\sigma$ is not essential for the geometry and has been set equal
to one. The space-time is regular for $r>0$ and, as with a previous special
case, the solution is also valid under a parity transformation $r\rightarrow
-r$, $\rho(r) = \sigma\, e^{-\frac{\sqrt{-\Lambda} r}{\sqrt{3}}}$ with the
manifold being now regular for $r<0$.

At the end we complete this analysis of particular solutions by checking the
case where $\Lambda=0$ and $\bar{K}=2$. This results in the solution where
both $F$ and $\rho$ being constants with $F(r)=-\rho(r)^{2}= - \rho_{0}^{2}$,
that lead to a regular Euclidean manifold when $\rho_{0}$ is real and
Lorentzian but a cosmological manifold (with $r$ as time) whenever $\rho_{0}$
is purely imaginary.

Let us note that all the above solutions of the $F=-\rho^{2}$ case admit six
Killing vector fields instead of just the four of the initial general line
element \eqref{Kilgen}, the extra two being
\begin{subequations}
\label{Killing}%
\begin{align}
&  \xi_{5}=\cot\theta\cos t\partial_{t}+\sin t\,\partial_{\theta}-\frac{\cos
t}{\sin\theta}\partial_{\phi}\\
&  \xi_{6}=\cot\theta\sin t\partial_{t}-\cos t\,\partial_{\theta}-\frac{\sin
t}{\sin\theta}\partial_{\phi}.
\end{align}

\section{The Klein-Gordon equation in the two regular Lorentzian families}

In this section a quite non-trivial property of the two regular Lorentzian
configurations of the Einstein nonlinear sigma model system will be discussed.
Namely, the Klein-Gordon equation on the metrics corresponding to Eqs.
(\ref{wormh1}), (\ref{wormh2}) and (\ref{metconst}) respectively is not only
separable but it is also integrable in terms of very well-known solvable potentials.

\subsection{The wormhole solution}

The Klein-Gordon equation
\end{subequations}
\begin{equation}
\left(  \square-m^{2}\right)  \Psi=0\ , \label{kg1}%
\end{equation}
on the metrics corresponding to Eqs. (\ref{wormh1}) and (\ref{wormh2}) (which
are allowed when $\bar{K}-8>0$) can be solved by separation of variables. That
is possible since equation (\ref{kg1}) is a linear equation, i.e. the Lie
symmetry vector $\Psi\partial_{\Psi}$ exists, and secondly the underlying
manifold which defines the box operator, is a locally rotational space-time
(LRS) and admits a four dimensional Killing algebra, the $A_{1}\oplus
\mathfrak{so}(3)~$\cite{tsaL}, where $A_{1}$ indicates the autonomous symmetry
vector $\partial_{t}$. Recall that the isometry vectors of the space-time
(\ref{metricnew}) generate Lie point symmetries for equation (\ref{kg1}), for
details see \cite{anpal}.

It is known that any Lie point symmetry $X=\partial_{I}+\alpha\Psi
\partial_{\Psi}$, is equivalent with the linear Lie-B\"acklund symmetry
$\bar{X}=\left(  \Psi_{,I}-\alpha\Psi\right)  \partial_{\Psi}$, and by the
definition of the symmetry vector that transform solutions into solutions we
have $\bar{X}\Psi=\left(  \gamma-\alpha\right)  \Psi$, or $\Psi_{,I}%
=\gamma\Psi$, that is, $\Psi=\Psi\left(  x^{i}\right)  e^{\gamma x^{I}}%
,~$solutions of that forms are called invariant solutions.

Hence, for equation (\ref{kg1}) with the use of the autonomous symmetry, the
rotation $\partial_{\phi}$, and the Casimir invariant of the $\mathfrak{so}%
\left(  3\right)  $ algebra, we find the following general form for an
invariant solution
\begin{equation}
\Psi(t,r,\theta,\phi)=U(r)Y(\theta)e^{i(\mu\,\phi+\omega\,t)},
\label{separansatz}%
\end{equation}
where $U\left(  r\right)  ,~Y\left(  \theta\right)  $, satisfies the following
set of linear second order differential equations%
\begin{align}
-\frac{1}{\sin\theta}\frac{d}{d\theta}\left(  \sin\theta\frac{d}{d\theta
}Y(\theta)\right)  +\left(  \frac{\omega\cos\theta-\mu}{\sin\theta}\right)
^{2}Y(\theta)  &  =\lambda Y(\theta)\ ,\label{wormkg1}\\
U^{\prime\prime}(r)+\sqrt{-3\,\Lambda}\tanh\left(  \sqrt{\frac{-\Lambda}{3}%
}\,r\right)  U^{\prime}(r)+\frac{16\,\Lambda\left(  \lambda\,\bar{K}%
-4\,\omega^{2}\right)  }{3(\bar{K}-8)\bar{K}\cosh^{2}\left(  \sqrt
{\frac{-\Lambda}{3}}\,r\right)  }U\left(  r\right)   &  =m^{2}U(r)\ ,
\label{wormkg2}%
\end{align}
where $\lambda$ is the separation constant. Equations \eqref{wormkg1},
\eqref{wormkg2} are linear and maximally symmetric, which implies that there
exists a transformation $\left\{  r,U,\theta,Y\right\}  \rightarrow\left\{
\bar{r},\bar{U},\bar{\theta},\bar{Y}\right\}  $, where they can be written in
the form of those of the free particle \cite{prince}; that is:
\begin{equation}
\frac{d^{2}\bar{U}}{d\bar{r}^{2}}=0~,~\frac{d^{2}\bar{Y}}{d\bar{\theta}^{2}%
}=0.
\end{equation}

Even though this property holds for any general form of the space-time
(\ref{metricnew}) - i.e. for arbitrary functions $F\left(  r\right)  $,
$\rho\left(  r\right)  $ - and such a transformation always exists, it is not
always possible to express the latter in a closed-form. However, as we see
bellow for the case at hand, the radial term of the Klein-Gordon equation
reduces to a well known linear equation in which the solution can be written
in closed-form.

In equation (\ref{wormkg2}) we do the following change of variable%
\begin{equation}
U\left(  r\right)  =\frac{\psi\left(  r\right)  }{\left[  \cosh\left(
\sqrt{\frac{-\Lambda}{3}}\,r\right)  \right]  ^{3/2}}\ , \label{wormkg3}%
\end{equation}
then we find
\begin{align}
\psi^{\prime\prime}+\frac{\Sigma}{\left[  \cosh\left(  \sqrt{\frac{-\Lambda
}{3}}\,r\right)  \right]  ^{2}}\psi &  =\left(  -\frac{3}{4}\Lambda
+m^{2}\right)  \psi\ ,\label{wormkg4}\\
\frac{\Lambda\left[  \left(  24+64\lambda-3\,\bar{K}\right)  \,\bar
{K}-256\omega^{2}\right]  }{12\bar{K}\left(  \bar{K}-8\right)  }  &
=\Sigma\ . \label{wormkg5}%
\end{align}

The angular equation (\ref{wormkg1}) reduces to the equation for the rotation
matrices which appear in the scattering problem of a monopole. In particular,
Eq. (\ref{wormkg1}) reduces to Eqs. (3.10) and (3.11) of \cite{Schwinger} with
the identifications%
\begin{align}
\lambda &  =j\left(  j+1\right)  -\left(  m`\right)  ^{2}\ ,\label{sc1}\\
\omega &  =m`\ ,\ \ \ \mu=\overline{m}\ ,\label{sc2}\\
j  &  \geq\left\vert m`\right\vert \ ,\ \ \left\vert \overline{m}\right\vert
\ \Rightarrow\ \lambda>0\ . \label{sc3}%
\end{align}
In the notation of \cite{Schwinger}, $m`$\ is the strength of the Dirac
monopole and can take only integers and half-integers values, $\overline{m}$
is the azimutal quantum number (restricted to be integer) while $j$ is the
eigenvalue of the total angular momentum. Therefore, just from the generic
properties of the rotation matrices, one gets that the frequency $\omega$
appearing in the invariant solution (\ref{separansatz}) is restricted to be
integer or half integer. Moreover, the separation constant $\lambda$ is
quantized as in Eq. (\ref{sc1}) and strictly positive.

On the other hand, the radial equation (\ref{wormkg4}) (with the coefficient
in Eq. (\ref{wormkg5})) corresponds to the Poschl-Teller potential (which is
one of the most famous exactly solvable potential in quantum mechanics: see,
for a review, \cite{postel}). Thus, one can extract a lot of information and
non-trivial constraints from well-known results on the Poschl-Teller
potential. First of all, the coefficient $\Sigma$ is required to be positive,
which implies the following requirement on the coupling constant $\bar{K}$ to
be large enough:%
\begin{equation}
\left(  24+64\lambda-3\,\bar{K}\right)  \,\bar{K} - 256\omega^{2}<0\ .
\label{pt1}%
\end{equation}
Secondly, one observes that the would-be eigenvalue $E^{2}$ of the Schrodinger
equation with Poschl-Teller potential (\ref{wormkg4}) is%
\[
E^{2}=-\left(  -\frac{3}{4}\Lambda+m^{2}\right)
\]
and therefore negative. Hence, the problem is solvable if the Poschl-Teller
potential in Eq. (\ref{wormkg4}) with the coefficient in Eq. (\ref{wormkg5})
admits bound states. Well known results on the Schrodinger equation with
Poschl-Teller potential tell us the number of bound states and the
corresponding eigenvalues. To proceed, let us first write Eq. (\ref{wormkg4})
with standard normalization changing the variable from $r$ to $x=\sqrt
{\frac{-\Lambda}{3}}\,r$:%
\begin{equation}
-\frac{d^{2}\psi}{dx^{2}}-\frac{3\Sigma}{\left(  -\Lambda\right)  \left[
\cosh\left(  x\right)  \right]  ^{2}}\psi=-\frac{\left(  -\frac{3}{4}%
\Lambda+m^{2}\right)  }{\left(  \frac{-\Lambda}{3}\right)  }\psi\ .
\label{pt2}%
\end{equation}
Then, let us write the coefficient of the potential as%
\begin{equation}
\frac{3\Sigma}{\left(  -\Lambda\right)  }=N\left(  N+1\right)  \ ,\ N>0\ .
\label{pt3}%
\end{equation}
The number $n_{B}$ of bound states in the Poschl-Teller potential is
\begin{equation}
n_{B}=\left\lfloor N\right\rfloor \ , \label{pt4}%
\end{equation}
where $\left\lfloor X\right\rfloor $ denotes the integer part of $X$. The
corresponding eigenvalues are given by the condition%
\begin{align}
\frac{\Gamma\left(  \zeta+1\right)  \Gamma\left(  \zeta\right)  }%
{\Gamma\left(  \zeta+N+1\right)  \Gamma\left(  \zeta-N\right)  }  &
=0\ ,\label{pt5}\\
\frac{\left(  -\frac{3}{4}\Lambda+m^{2}\right)  }{\left(  \frac{-\Lambda}%
{3}\right)  }  &  =-\zeta^{2}\ , \label{pt6}%
\end{align}
where $\Gamma$ is the Euler gamma-function. Therefore, assuming $\zeta$\ to be
positive, the eigenvalues are determined by the the condition that $\zeta-N$
is a negative integer:%
\begin{equation}
-\left(  \zeta-N\right)  \in%
%TCIMACRO{\U{2115} }%
%BeginExpansion
\mathbb{N}
%EndExpansion
\ ,\ \ \zeta-N<0\ . \label{pt7}%
\end{equation}
In particular, the allowed values for $m^{2}$ in the Klein-Gordon equation are
fixed by the quantization condition in Eq. (\ref{pt7}). Finally, in order to
have at least one bound state in the radial Schrodinger problem one must
require that%
\begin{equation}
\frac{3\Sigma}{\left(  -\Lambda\right)  }=N\left(  N+1\right)  \geq2\ .
\label{pt8}%
\end{equation}

\subsection{The cylindrical solution}

The Klein-Gordon equation
\[
\left(  \square-m^{2}\right)  \Psi=0\ ,
\]
on the metrics corresponding to Eq. (\ref{metconst}) (which are allowed when
$\bar{K}=8$ and $\Lambda=0$) and as before is invariant under the same group
of transformations, hence it can be solved with the method of separation of
variables. Therefore, the group invariant ansatz (\ref{separansatz}),
substituted in the Klein-Gordon equation results in the following system
\begin{align}
-\frac{1}{\sin\theta}\frac{d}{d\theta}\left(  \sin\theta\frac{d}{d\theta
}Y(\theta)\right)  +\left(  \frac{\omega\cos\theta-\mu}{\sin\theta}\right)
^{2}Y(\theta)  &  =\lambda Y(\theta)\ ,\label{cyl1}\\
U^{\prime\prime}(r)-\left(  m^{2}+\frac{\lambda-2\,\omega^{2}}{4\,\rho_{0}%
^{2}}\right)  U(r)  &  =0\ , \label{cyl2}%
\end{align}
where $\lambda$ is the separation constant.

In the present case, the angular equation is the same as in the previous
subsection and can be analyzed following \cite{Schwinger}. In particular, the
solvability of the problem once again requires the quantization conditions in
Eqs. (\ref{sc1}) and (\ref{sc2}) (which implies that $\lambda>0$) for $C$ and
$\omega$. In this case, the radial equation is simpler and one gets the
typical behavior of a free massive field when%
\begin{equation}
m^{2}+\frac{\lambda-2\,\omega^{2}}{4\,\rho_{0}^{2}}<0\ , \label{cyl3}%
\end{equation}
in which case one gets periodic solutions along the axis of the cylinder. On
the other hand, when%
\[
m^{2}+\frac{\lambda-2\,\omega^{2}}{4\,\rho_{0}^{2}}>0\ ,
\]
one gets solutions which are unbounded on, at least, one side of the cylinder
(either $r\rightarrow-\infty$ or $r\rightarrow+\infty$). Therefore, the
inequality in Eq. (\ref{cyl3}) must be satisfied.

In conclusions, in both cases the solvability of the Klein-Gordon equation
implies strong constraints on the parameters of the theory.

\section{Conclusions and perspectives}

Exact, regular and topologically non-trivial\ configurations of the
Einstein-nonlinear sigma model system in (3+1) dimensions have been
constructed. The ansatz is the $SU(2)$ generalization of the usual boson star
ansatz for charged $U(1)$ fields. Moreover, the $SU(2)$ configuration cannot
be deformed continuously to the trivial Pions vacuum as it possesses a
non-trivial winding number. Due to the fact that, with the chosen ansatz for
the metric, the nonlinear sigma model satisfies identically the corresponding
field equations, the full coupled Einstein non-linear sigma model system can
be successfully reduced to a problem that involves a single second order ODE
(our master equation). When the cosmological constant vanishes, such master
equation can be further reduced to an Abel equation. This is quite a technical
achievement in itself and it can be used as a very convenient starting point
by researchers working in this field. In particular, such master equation can
reveal whether or not there are black holes in this sector of the theory. We
hope to come back on this issue in a future publication.

Such master equation can describe both Lorentzian and Euclidean space-times.
Here we have described in more details the properties of regular and smooth
Lorentzian solutions. However, the regular Euclidean solutions, due to their
non-trivial topological properties could be interpreted as instantons of the
theory. We hope to come back on this interesting issue in a future publication.

The most interesting regular Lorentzian configurations correspond to a
stationary traversable wormhole with NUT parameter and a (3+1)-dimensional
cylinder whose (2+1)-dimensional sections are Lorentzian squashed spheres.

The use of the theory of group invariant transformations reveals that the
Klein-Gordon equations in these two families of space-times are not only
separable but the reduced equation lead to well known quantum systems where
the solution can be written in terms of special functions. The angular
equation reduces to the equation for the Jacobi polynomial (typical of the
Schrodinger problem in the field of a Dirac monopole) while the effective
potential appearing in the radial equation belongs to the Poschl-Teller
family. The solvability requirement for the Poschl-Teller equation determines
a quantization condition for the parameters of the theory.

In the case in which the cosmological constant vanishes, the solutions of the
master equation (\ref{sym.03}) that we have found are asymptotic solutions,
that is, asymptotically, the general solution will be described by one of
these solutions. The reason for that is that the families of regular
configurations that we have found are fixed points for equation (\ref{sym.03}%
). We hope to extend the present analysis in the case of the Euclidean
configurations in a future work.

\begin{acknowledgments}
The authors want to warmly thank E. Ayon-Beato and J. Zanelli for fruitful
discussions. This work has been funded by the Fondecyt grants 1160137 (FC),
3150016 (ND) and 3160121 (AP). The Centro de Estudios Cient\'{\i}ficos (CECs)
is funded by the Chilean Government through the Centers of Excellence Base
Financing Program of Conicyt. AP thanks the Durban University of Technology
for the hospitality provided while part of this work was performed.
\end{acknowledgments}

\bigskip

\end{document}